\documentstyle[aps,prl,floats]{revtex}

\begin{document}
\input epsf
\draft
\renewcommand{\floatpagefraction}{0.99}
\renewcommand{\topfraction}{0.99}
\twocolumn[\hsize\textwidth\columnwidth\hsize\csname@twocolumnfalse%
\endcsname
\title{  {\rm\small\hfill Surf. Sci. (in press)}\\
Surface Core-Level Shifts at an oxygen-rich Ru Surface:\\ O/Ru(0001) vs. RuO${}_2$(110)}
\author{Karsten Reuter and Matthias Scheffler}
\address{Fritz-Haber-Institut der Max-Planck-Gesellschaft, Faradayweg
4-6, D-14195 Berlin-Dahlem, Germany}
\date{Received 12 April 2001}
\maketitle

\begin{abstract}
We present density-functional theory calculations of Ru $3d$ and O $1s$
surface core-level shifts (SCLSs) at an oxygen-rich Ru(0001) surface,
namely for the O$(1 \times 1)$/Ru(0001) chemisorption phase and for two
surface terminations of fully oxidized RuO${}_2$(110). Including
final-state effects, the computed SCLSs can be employed for the analysis of
experimental X-ray photoelectron spectroscopy (XPS) data enabling a detailed
study of the oxidation behaviour of the Ru(0001) surface. We show that
certain peaks can be used as a fingerprint for the existence of the various
phases and propose that the long disputed satellite peak in RuO${}_2$(110)
XPS data originates from a hitherto unaccounted surface termination.

\vspace*{0.5cm}
\end{abstract}
\hfill {\quad}
]

\section{Introduction} 
In addition to the long-standing interest in RuO${}_2$ as a technologically
relevant metallic oxide {\em per se}, this material has recently also
received a lot of attention as the end product in the oxidation sequence of
ruthenium surfaces \cite{boettcher97,boettcher00,over00,kim01a}. Epitaxially
well oriented, incommensurate RuO${}_2$(110) domains were identified on
Ru(0001) after heavy oxygen exposure and related to the enhanced catalytic
activity towards CO oxidation \cite{over00,kim01b}, observed at Ru under
high oxygen partial pressure \cite{boettcher97,peden86}. Subsequently, an
atomistic mechanism for this surface-oxide formation was proposed
\cite{reuter01}, in which after the completion of the full monolayer
O$(1 \times 1)$ chemisorption phase on the surface, O penetrates into the
sample and clusters in islands between the first and second metallic layer,
locally decoupling a O-Ru-O trilayer from the underlying substrate. 
The ongoing oxidation results in more and more of these trilayers, which
finally unfold into the more open rutile structure giving rise to (110)
oriented RuO${}_2$ patches on the surface.

Depending on the O partial pressure, a real Ru(0001) surface will therefore
exhibit a rather complex morphology, ranging from chemisorbed O overlayers
to sub-surface O and RuO${}_2$(110) patches with different surface terminations.
Given this low degree of structural order, X-ray photoelectron spectroscopy
(XPS) appears as a well suited tool to study the oxidation process in
detail. However, the complex core-level spectra to be expected from such
a surface with a manyfold of peaks due to coexisting domains will render the
interpretation of the obtained data not unambiguous. This is even aggravated
by the fact, that already for the RuO${}_2$(110) surface alone
the origin of the peaks has been discussed controversially
\cite{kim74,lewerenz83,kotz83,grady84,atanasoska88,chan97,cox86,kim97}.

To provide a theoretical guidance for the experimental data analysis, we thus
calculated all Ru $3d$ and O $1s$ surface core-level shifts (SCLSs) for the
individual, already identified domains on an oxygen-rich Ru(0001) surface,
namely the O$(1 \times 1)$ chemisorbed phase, as well as two different surface
terminations of RuO${}_2$(110). With this, we will show that certain peaks may
be used as a fingerprint for the existence of one or the other domain.

\section{Theoretical}

The SCLSs are obtained from density-functional theory (DFT) calculations
within the generalized gradient
approximation (GGA) of the exchange-correlation functional \cite{perdew96}.
The full-potential linear augmented plane wave method (FP-LAPW) 
\cite{blaha99,kohler96,petersen00} is used to solve the Kohn-Sham equations.
We model the O$(1 \times 1)$/Ru(0001) surface with a six layer slab, adsorbing
O on both sides to preserve mirror symmetry. Likewise, the RuO${}_2$(110)
surface is represented by a symmetric slab consisting of 3 rutile 
O-(RuO)-O trilayers. A vacuum region of $\approx$11{\AA} was employed to
decouple the surfaces of consecutive slabs in the supercell approach. The
resulting, fully relaxed geometries are in very good agreement with previous
LEED and DFT studies \cite{over00,stampfl96a}.

The FP-LAPW basis set is taken as follows: $R_{\rm{MT}}^{\rm{Ru}}=$1.8 bohr,
$R_{\rm{MT}}^{\rm{O}}=$1.3 bohr, wave function expansion inside the muffin
tins up to $l_{\rm{max}}^{\rm{wf}} = 12$, potential expansion up to
$l_{\rm{max}}^{\rm{pot}} = 4$. The Brillouin zone integration for the
O$(1 \times 1)$/Ru(0001) and RuO${}_2$(110) was performed using a
$(12 \times 12 \times 1)$ and $(5 \times 10 \times 1)$ Monkhorst-Pack grid
with 19 and 15 {\bf k}-points in the irreducible part respectively.
The energy cutoff for the plane wave representation in the interstitial
region between the muffin tin spheres was 17 Ry for the wave functions and
169 Ry for the potential.

The SCLS, $\Delta_{\rm{SCLS}}$, is defined as the difference in energy which is
needed to remove a core electron either from a surface or from a bulk atom
\cite{spanjaard85}. We calculate this quantity both in the initial-state
approximation, as well as taking final-state effects into account. In the
prior approximation, the SCLS is simply due to the variation of the orbital
eigenenergies before the excitation of the core electron. Yet, the total, measurable
SCLS involves an additional component due to the screening contribution from
the valence electrons in response to the created core hole. We obtain this
total shift, $\Delta_{\rm{SCLS}}^{\rm total}$, via the Slater-Janak transition
state approach of evaluating total energy differences \cite{janak78} using
impurity type calculations where an ionized atom is once placed at the surface
and once in the bulk. In order to decouple these ionized atoms from each
other, we surround them with neighbours possessing the normal core configuration
by performing the calculation in a $(2 \times 2)$ and a $(2 \times 1)$
supercell for Ru(0001) and RuO${}_2$(110) respectively. Having calculated
both the total and the initial-state SCLS allows us then to single out the
final-state effect from their respective difference. The procedure and numerical
accuracy of both types of calculations has been described in detail in a recent
publication addressing the SCLSs of all ordered O adlayers on Ru(0001)
\cite{lizzit01}: Comparing with a large set of experimental high-resolution
core-level shift data we achieved consistent agreement to within
$\approx \pm 0.05$ eV, confirming the high accuracy and {\em predictive
nature} of our calculations.

However, we immediately stress that we do not expect such a precision when
comparing metallic with oxidic core-levels. SCLSs are obtained from the
difference between a surface and a bulk calculation and thus benefit from
error cancelation, if both calculations are done on the same substrate
(which eletronic structure may be equally well described by DFT-GGA) and
within the same supercell. While this is obviously not the case for Ru
vs. RuO${}_2$, notice that the focus of the present work is also not the
explanation of a highly accurate existing data set on a very well defined
sample. Rather, we intend to provide a fingerprint guidance for the XPS
trends on an oxidizing Ru surface, where none of the conclusions drawn
will be affected by a likely $\approx \pm 0.2$ eV inaccuracy between the
position of the Ru and RuO${}_2$ bulk peaks.

\section{O$(1 \times 1)$/Ru(0001) and RuO${}_2$(110) surface structure}

\begin{figure}
       \epsfxsize=0.35\textwidth \centerline{\epsfbox{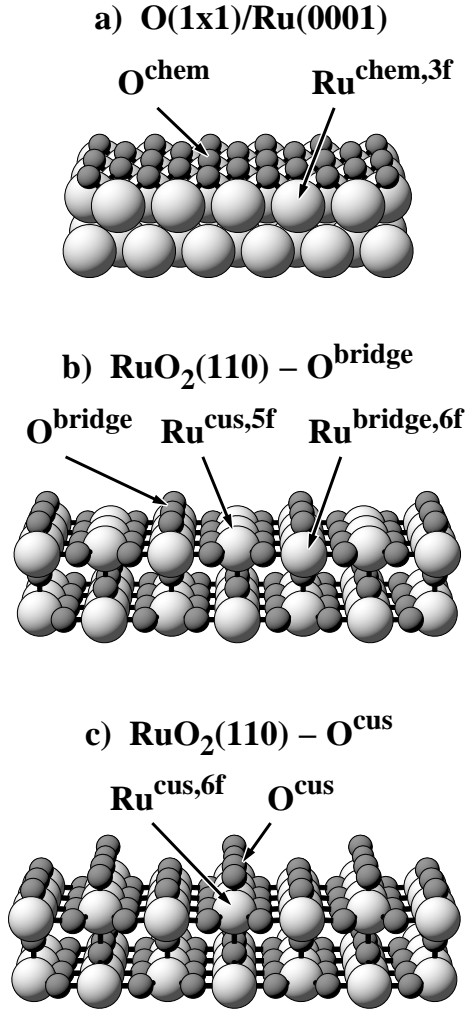}}
       \caption{Surface structures of domains on O-rich Ru(0001):
        a) Chemisorbed O$(1 \times 1)$/Ru(0001) with threefold
        O${}^{{\rm chem}}$ coordinated Ru${}^{{\rm chem, 3f}}$ surface
        atoms. b) Stoichiometric RuO${}_2$(110)-O${}^{{\rm bridge}}$
        termination with five-, six- and twofold coordinated
        Ru${}^{{\rm cus, 5f}}$, Ru${}^{{\rm bridge, 6f}}$ and O${}^{{\rm bridge}}$
        atoms respectively. c) High-pressure RuO${}_2$(110)-O${}^{{\rm cus}}$
        termination, where additional O${}^{{\rm cus}}$ atoms sit atop the
        formerly undercoordinated Ru${}^{{\rm cus, 6f}}$ atoms (Ru = large,
        light spheres, O = small, dark spheres).
        \label{surfstructures}}
\end{figure}

The initial oxygen chemisorption on the Ru(0001) surface proceeds via four
ordered adlayer structures, in which with increasing coverage O consecutively
occupies the four available hcp sites of a $(2 \times 2)$ unit cell
\cite{stampfl96b,menzel99}. Sub-surface O penetration does not begin
until the full O$(1 \times 1)$ monolayer has been completed 
\cite{reuter01,stampfl96a,boettcher99a}, which is why only the latter
adphase may coexist with already fully oxidized patches under O-rich 
conditions \cite{kim01a}. In the O$(1 \times 1)$, every Ru surface atom,
Ru${}^{{\rm chem, 3f}}$, is coordinated to three chemisorbed oxygens,
O${}^{{\rm chem}}$, with a bondlength of 1.97 {\AA}, cf. Fig.
\ref{surfstructures}a. Note, that we will use a nomenclature for the
surface Ru atoms, where apart from a site specific characterization (e.g.
chem for the chemisorption phase) also the number of direct O
neighbours (e.g. 3f for 3-fold coordination) is indicated. Vice versa,
we indicate for the surface O atoms to which specific site they bind (e.g.
O${}^{{\rm chem}}$ binds to the Ru${}^{{\rm chem, 3f}}$ atoms). In the
rutile bulk structure of RuO${}_2$ \cite{sorantin92} the coordination
is much higher compared to the chemisorption phase, albeit at a very similar
bondlength: Every Ru atom is surrounded by an octahedron of six oxygens with
O-Ru bondlengths of 2.00 {\AA} and 1.96 {\AA} to the four basal and two apical
oxygens respectively. On the other hand, the Ru coordination of O atoms in
the chemisorbed state and in the bulk oxide structure is both threefold.

If the rutile RuO${}_2$ is cut along the (110) plane, three distinct
surface terminations are possible. The stoichiometric termination shown
in Fig. \ref{surfstructures}b exhibits two types of atoms with truncated
bonds, namely fivefold coordinated Ru${}^{{\rm cus, 5f}}$ and twofold 
coordinated O${}^{{\rm bridge}}$ atoms. As a result, both atoms relax
inwards, which e.g. translates into a shortened bondlength of 1.91
{\AA} between O${}^{{\rm bridge}}$ and the underlying sixfold coordinated
Ru${}^{{\rm bridge, 6f}}$ atoms. This termination (henceforth referred to as
O${}^{{\rm bridge}}$ termination) is traditionally believed to be the most
stable one for (110) surfaces of all rutile-structured metal oxides
\cite{henrich94}, and in ultrahigh vacuum (UHV) it was also found on the oxide
patches that had formed on the Ru(0001) surface \cite{over00}. However, in
recent calculations determining the lowest energy structure in equilibrium
with a given environment we found that this termination is only stable at
a low O chemical potential \cite{reuter01b}. Under more O-rich conditions a
different termination (henceforth referred to as O${}^{{\rm cus}}$ termination)
is stabilized, in which terminal oxygen atoms, O${}^{{\rm cus}}$, occupy
sites atop of the formerly coordinatively unsaturated Ru${}^{{\rm cus, 5f}}$
atoms as shown in Fig. \ref{surfstructures}c. Although this achieves the
bulk-like sixfold O coordination for the now Ru${}^{{\rm cus, 6f}}$ atoms, the
O${}^{{\rm cus}}$-Ru${}^{{\rm cus, 6f}}$ bondlength is due to the
singly-bonded atop site with 1.70 {\AA} significantly shorter than all
aforementioned bondlengths in the O/Ru system.

By post-dosing the RuO${}_2$(110) surface with additional O at room
temperature this high-pressure termination of RuO${}_2$(110) has recently
also been created and characterized in UHV intentionally \cite{kim01c,fan01}.
However, depending on the details of the experimental preparation we are
convinced that this high-pressure termination was probably also present
in a number of previous studies addressing RuO${}_2$(110), in particular
the ones where the oxide was treated under O-rich conditions without a final
annealing step \cite{boettcher97,boettcher00,kim97,boettcher99a}. We finally
note in passing, that our calculations show that the third possibility
of terminating a RuO${}_2$(110) crystal with a mixed (RuO) layer exhibited
at the surface is never realized in the range of possible O chemical
potentials.

\section{Results}

\subsection{Surface Core-Level Shifts for RuO${}_2$(110)} 

\begin{table} 
\caption{\label{sclstable}
Calculated Ru $3d$ and O $1s$ SCLSs on RuO${}_2$(110). Shown are the
total shifts, as well as their decomposition into screening and initial
state parts:  $\Delta_{\rm{SCLS}}^{\rm{total}} = \Delta^{\rm{screen}} + 
\Delta_{\rm{SCLS}}^{\rm{initial}}$.}  

\begin{tabular}{ll | r | rr}
\multicolumn{5}{c}{\rule[-1.9mm]{0cm}{0.5cm}Ru $3d$ SCLSs in eV} \\ \hline
Termination:          &                            & Total        & Screening  & Initial    \\ \hline 
O${}^{{\rm bridge}}$  & Ru${}^{{\rm bridge, 6f}}$  & +0.29        & -0.15      & +0.44      \\
                      & Ru${}^{{\rm cus, 5f}}$     & -0.16        & -0.02      & -0.14      \\[0.3cm]
O${}^{{\rm cus}}$     & Ru${}^{{\rm bridge, 6f}}$  & +0.40        & -0.14      & +0.54      \\
                      & Ru${}^{{\rm cus, 6f}}$     & +1.37        & +0.07      & +1.30      \\[0.3cm] \hline
\multicolumn{5}{c}{\rule[-1.9mm]{0cm}{0.6cm}O $1s$ SCLSs in eV} \\ \hline
Termination:          &                            & Total        & Screening  & Initial    \\ \hline 
O${}^{{\rm bridge}}$  & O${}^{{\rm bridge}}$       & -0.87        & -0.68      & -0.19      \\[0.3cm]
O${}^{{\rm cus}}$     & O${}^{{\rm bridge}}$       & -0.97        & -0.77      & -0.20      \\
                      & O${}^{{\rm cus}}$          & -0.79        & -0.98      & +0.19      \\
\end{tabular} 
\end{table} 

Aiming at a fingerprint guidance for XPS experiments, we concentrate on the
discussion of the photoemission from those atoms, for which rather large
SCLSs are expected due to a significantly changed local environment with respect to
the bulk situation, cf. Table \ref{sclstable} and Fig. \ref{surfstructures}.
From an inspection of the initial-state SCLSs we estimate that the peaks
due to all other atoms, which are located in deeper surface layers, will
lie closer than $\approx \pm$ 0.2 eV around the respective RuO${}_2$ bulk
peak. Note, that in our sign convention a positive SCLS indicates a shift
towards higher binding energies, i.e. the core-level moves away from the 
Fermi level. Similarly, a positive screening contribution to the total
shift occurs, if the created core hole is less screened at the surface
than in the bulk.

Concentrating first on the Ru $3d$ SCLSs in the stoichiometric
RuO${}_2$(110)-O${}^{\rm bridge}$ termination we find only rather modest
shifts for both the Ru${}^{\rm bridge, 6f}$ and the Ru${}^{\rm cus, 5f}$ atoms.
While this is not very surprising for the former atom, which possesses its
sixfold bulk-like O coordination (albeit with a reduced bondlength to the
O${}^{\rm bridge}$ atoms), one could have imagined a larger shift for the
Ru${}^{\rm cus, 5f}$ atoms, which after all lack their atop apical oxygen
neighbour. However, these atoms are considerably relaxed inwards, reinforcing
the backbond to the underlying second apical O at a reduced bondlength of 1.88
{\AA}. Thus, the bond truncation and backbond strengthening seem to balance
up, leading in total to an almost bulk-like situation and in turn to a very
small SCLS.

The final-state contribution for both atoms is negative, indicative
of a more efficient screening at the surface. A similar screening behaviour was
already found for all Ru surface atoms in the O adlayer phases on metallic
Ru(0001) and explained by an enhanced $d$ density of states (DOS) around the
Fermi level \cite{lizzit01}: Upon core excitation, the $4d$-band shifts to
lower energies, causing a valence electron from the Fermi reservoir to restore
local charge neutrality by filling up formerly unoccupied $d$-states.
In turn, if the $d$-DOS at and above the Fermi level is higher for a
surface atom than for a bulk atom, a more efficient screening results.
Inspecting the $d$-DOS of both Ru${}^{\rm bridge, 6f}$ and Ru${}^{\rm cus, 5f}$, we
indeed find again such an enhancement, confirming that not only in the
Ru metal, but also in its oxide the final-state effect is mainly due to
intra-atomic $d$-electron screening.

In the RuO${}_2$(110)-O${}^{\rm cus}$ termination, the Ru${}^{\rm bridge, 6f}$ SCLS 
remains almost unchanged, which is intelligible as the addition of the terminal
O${}^{\rm cus}$ atoms does not directly influence the nearest-neighbour
coordination of this Ru surface atom, cf. Fig. \ref{surfstructures}c. However,
the Ru${}^{\rm cus, 6f}$ atoms are of course significantly affected, exhibiting
now a large shift of +1.37 eV towards higher binding energies. As apparent
from Table \ref{sclstable} this shift is primarily an initial-state effect,
which we attribute to the very short bondlength of 1.70 {\AA} formed to the
atop O${}^{\rm cus}$ atoms. Yet, also the Ru${}^{\rm cus, 6f}$ $d$-DOS at the
Fermi level is lowered so strongly by the new bond, that we additionally
observe a sign reversal in the screening contribution. Reflecting a reduced
screening capability at the surface the final-state effect points thus in
the same direction as the large initial-state shift, further augmenting the
total SCLS. Note, that we had previously observed such a change in the
screening behaviour also at high coverage O adlayers on Rh(111), where the
oxygen-metal interaction likewise reduced the $d$-DOS at the Fermi level below
the corresponding bulk value \cite{ganduglia01}. The resulting very large
SCLS for the Ru${}^{\rm cus, 6f}$ atoms in this hitherto unaccounted high-pressure
termination is related to a long standing controversy concerning
the interpretation of XPS data on RuO${}_2$(110), which detailed discussion
we defer to the following section.

\begin{figure}
       \epsfxsize=0.47\textwidth \centerline{\epsfbox{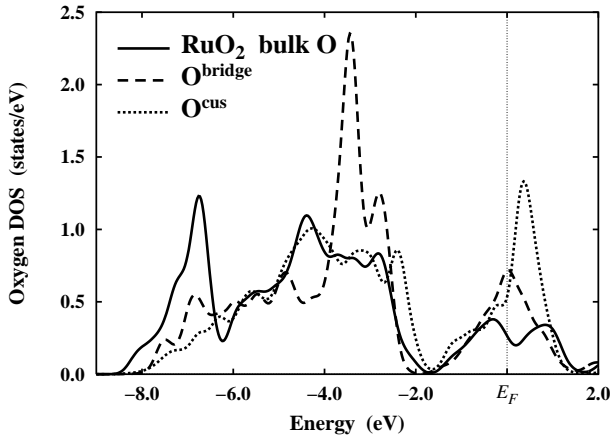}}
       \caption{Calculated DOS for bulk O in RuO${}_2$ (solid line), as
        well as for O${}^{\rm bridge}$ and O${}^{\rm cus}$ atoms
        (dashed and dotted lines respectively) in the RuO${}_2$(110)-O${}^{\rm
        cus}$ termination. The O${}^{\rm bridge}$ DOS in the
        RuO${}_2$(110)-O${}^{\rm bridge}$ termination is almost identical
        to the one shown here. The energy zero is at the Fermi level.
        \label{dos}}
\end{figure}

Turning our attention to the O $1s$ shifts, we immediately recognize that
contrary to the situation for the Ru $3d$'s, the total SCLSs are now
predominantly determined by a large, negative screening contribution.
The initial-state shifts on the other hand are almost bulk-like, with
O${}^{\rm bridge}$ atoms showing very similar values in both terminations.
The strongly enhanced screening at the surface is due to dangling-bond
type states on both surface oxygens as reflected in the DOS shown in
Fig. \ref{dos}. These oxygen-metal states, which are concentrated on
O${}^{\rm bridge}$ and O${}^{\rm cus}$ and their respective directly
bonded Ru${}^{\rm bridge, 6f}$ and Ru${}^{\rm cus, 6f}$ neighbours, fall in the
energy region -8.0 eV $< E <$ -2.0 eV in bulk RuO${}_2$ \cite{sorantin92},
but are shifted towards the Fermi level due to the bond truncation at
the surface. There they strongly enhance the DOS and thus induce the
large negative screening contribution to the O $1s$ SCLSs.

\subsection{The satellite peak discussion}

Motivated by the widespread use of RuO${}_2$ as catalyst for electrochemical,
as well as organic and inorganic processes, a large number of studies has
already applied low-resolution XPS to elucidate the oxide's surface structure
and composition (\cite{kim74,lewerenz83,kotz83,grady84,atanasoska88,chan97,cox86,kim97}
and references therein). Even at the resolution of a Mg/Al X-ray source the
Ru $3d$ spectrum of RuO${}_2$(110) clearly shows an additional rather broad peak
at about $+ 1.7 \pm 0.1$ eV to the higher binding energy side of each of the
primary 3/2 and 5/2 components \cite{atanasoska88,cox86,kim97}. Under the
assumption that the large shift of this peak indicates a significant deviation
of the local environment of the emitting atom with respect to the bulk phase,
the satellite was at first assigned to the presence of a higher oxidation
state of Ru at the surface, namely Ru${}^{6+}$ in a RuO${}_3$ type oxide
\cite{kim74,lewerenz83,kotz83,grady84,atanasoska88}. Concomitantly, this
interpretation resulted even in the inclusion of a Ru $3d_{5/2}$ binding energy
for RuO${}_3$ in a common XPS handbook \cite{moulder92}. However, more
recent work has questioned the existence of this conjectured surface
oxide, which is not known to be stable as a bulk phase \cite{cox86,kim97}.
In particular, a recent analysis of XPD azimuthal scans
by Kim {\em et al.} showed that the photoelectrons in the satellite peaks
originate from a rutile-type environment \cite{kim97}, which is an unlikely
structure for the presumed RuO${}_3$.

Alternatively, Cox {\em et al.} \cite{cox86} attributed the satellite peak 
to final-state screening effects. As outlined in the previous section, the
core hole created in the photoemission process pulls down formerly unoccupied
valence band DOS below the Fermi level. The photoemission spectrum may then
exhibit a multiple peak structure, often called the Kotani-Toyozawa effect
\cite{kotani73}: The high kinetic energy peak reflects the possibility that
electron transfer into the shifted DOS happens fast enough to impart the
corresponding screening energy onto the emitted photoelectron (this
corresponds to the fully-screened results of our calculations). And the
low-energy peak reflects the possibility that the shifted DOS remains
unoccupied on the time scale of the emission process. Of course, in
addition there is also the possibility of even lower kinetic energy
structures than this ``unscreened peak'' because the emitted 
photoelectron may also loose energy, e.g. by creating surface plasmons.
The Kotani-Toyozawa effect has been observed in photoemission from
systems with highly localized $f$ or $3d$ valence band states
\cite{kotani99}. However, the good overlap and spatial extension of the 
$4d$ wavefunctions in RuO${_2}$ (the width of the shifted DOS peak is 
about 1 eV) suggests that the screening dynamics will be fast. Thus,
the probability (i.e., the intensity) of this unscreened final-state
peak should be low. Furthermore, our calculations show that the
screening energy is small for emission from Ru core levels, cf. Table
\ref{sclstable}. Thus, even if a Kotani-Toyozawa peak existed at the
higher binding energy side, it would be shifted by less than 0.2 eV with
respect to the peak due to the fully-screened final-state, which is not
enough to account for the well separated satellite seen in the experiments.

On the other hand, for photoemission from the oxygen atoms, the dynamics
of screening could be interesting: The nodeless O $2p$ states are very
localized and the calculated screening energy is large. Consequently,
the screened and unscreened surface peaks of the photoemission spectrum
would be well separated. While the latter overlap (at least partly)
with the bulk peak, the fully-screened final-state peaks are noticeably
shifted to higher kinetic energy. Hence, high-resolution O $1s$
photoemission focusing on the low binding energy side of the bulk peak
(i.e. the peak shape) could provide important information on the many-body
dynamics of the photoemission process of this system.

Coming back to the Ru $3d$ satellite peak issue, we suggest in light
of the SCLS analysis presented in the previous section, that it is neither
due to RuO${}_3$, nor due to unscreened emission, but receives a signal
from the Ru${}^{\rm cus, 6f}$ atoms in the RuO${}_2$(110)-O${}^{\rm cus}$
high-pressure termination. Although the calculated large shift of +1.37 eV
agrees at first sight only semiquantitatively with the experimentally reported
$1.7 \pm 0.1$ eV (cf. the compilation of measured XPS data in Ref. \cite{chan97}),
one has to keep in mind that all these studies were done with low-resolution
XPS on rather ill-defined samples, the majority even on polycristalline material.
In particular, none of the studies was performed on single-crystal
RuO${}_2$(110), but always involving either oxidized Ru or grown RuO${}_2$
thin films, both bearing a certain likelihood for the presence of unoxidized
Ru remnants on the surface. As we will show below, the metallic Ru bulk peak
lies to the lower binding energy side of the RuO${}_2$ bulk peak, with all
surface peaks due to Ru coordinated to chemisorbed oxygen in
between. Unresolved, these peaks will lead to an erroneous RuO${}_2$ bulk peak
determination at too low a binding energy and in turn to an overestimation of
the total shift to the satellite peak, possibly explaining the 0.3 eV
difference to our result.

Asides, the assignment of the satellite peak as a fingerprint for the
hitherto simply not conceived high-pressure RuO${}_2$(110)-O${}^{\rm cus}$
termination would be fully compatible with the reported experimental sample
preparations, which unanimously involve highly oxidizing conditions.
After the transfer to UHV the O${}^{\rm cus}$ atoms are stable up to about
450 K \cite{boettcher99b}, so that the high-pressure termination is most
likely frozen in, if not a final annealing step is performed as e.g. in
the LEED work by Over and coworkers \cite{over00,kim01a,kim01b}. The
Ru${}^{\rm cus, 6f}$ atoms in the RuO${}_2$(110)-O${}^{\rm cus}$
termination are also obviously situated in a rutile-type environment,
thus explaining the aforementioned XPD results of Kim {\em et al.}
\cite{kim97}. Finally, we notice that our present model does not exclude
the possibility that a many-body peak, e.g. due to a surface plasmon
loss, falls in the same energy region, then being the predominant source
of the satellite peak. At present, we can only state that the screened
emission from Ru${}^{\rm cus, 6f}$ atoms - resolved or unresolved - contributes
to the signal at the corresponding energy. Whether this is the sole
explanation for the satellite peak or not, can only be determined by
a high-resolution XPS study on single-crystalline RuO${}_2$(110),
which we hope to encourage with the present work.

\subsection{XPS fingerprinting of the oxidation sequence}

\begin{figure}
       \epsfxsize=0.50\textwidth \centerline{\epsfbox{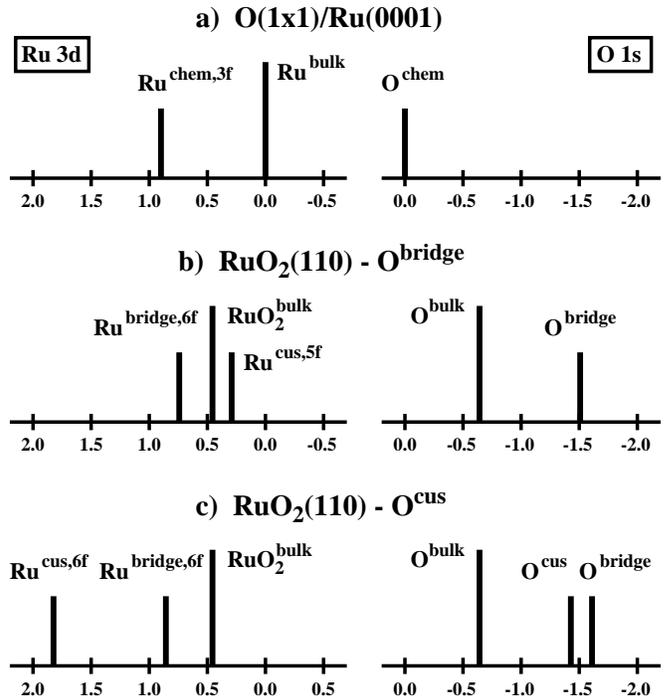}}
       \caption{Computed fully-screened Ru $3d$ and O$1s$ SCLSs for the three
        domain types on an oxygen-rich Ru(0001) surface. Ru in the
        metallic bulk and O in the O$(1 \times 1)$/Ru(0001) phase are
        used as zero reference. Units are eV.
        \label{sclstrend}}
\end{figure}

Having completed the analysis of the oxidic SCLSs, we proceed to compare
them with the peaks that arise from metallic Ru surface atoms coordinated
to chemisorbed O. Since our intention focuses on describing the XPS trends 
during the oxidation of a Ru(0001) surface, we now reference all Ru $3d$
and O $1s$ SCLSs with respect to the initially present peaks due to bulk
metallic Ru and chemisorbed O${}^{{\rm chem, 3f}}$ in the O$(1 \times 1)$/Ru(0001)
phase respectively. Again we stress that the shift between metallic and
oxidic bulk peaks cannot be determined as accurately as the individual
SCLSs on the respective surfaces and that we aim now more at the trends
than at ironclad numbers. On the other hand, note that on a {\em qualitative
level} our results even describe already the existing data from polycrystalline
samples \cite{kim74,lewerenz83,chan97}, Ru(0001) being the most stable
surface orientation.

In our previous work addressing the ordered O overlayers on Ru(0001)
we had shown that the progressive O chemisorption leads to a Ru $3d$
shift towards higher binding energies, which scales linearly with the
number of direct O neighbours coordinated to the respective Ru surface
atoms \cite{lizzit01}. For the final O$(1 \times 1)$/Ru(0001), which in
turn is precedent to and coexisting with already fully oxidized patches
on an O-rich Ru(0001) surface \cite{kim01a}, this results in a SCLS of
+0.90 eV for
the Ru${}^{\rm chem, 3f}$ atoms as shown in Fig. \ref{sclstrend}a (cf. with
the atomic structure displayed in Fig. \ref{surfstructures}a). On the
contrary, we find the O $1s$ level almost constant to within +0.18
eV in all four adlayer structures, indicating that the always threefold
coordinated O${}^{\rm chem, 3f}$ remains essentially in the same chemical
state apart from a slightly increased repulsion due to the more and more
close packing of the electronegative adsorbates in the higher coverage
phases.

With the given accuracy caveat, we determine the Ru $3d$ peak due to
bulk RuO${}_2$ at +0.46 eV on the higher binding energy side of
the metallic Ru peak, which compares well with the polycristalline
XPS literature value of $+0.7 \pm 0.1$ eV \cite{chan97}. As apparent in
Fig. \ref{sclstrend}b all Ru $3d$ peaks due to the UHV
RuO${}_2$(110)-O${}^{\rm bridge}$ termination will thus fall between the
Ru${}^{\rm chem, 3f}$ peak due to the Ru surface atoms in the just
described O$(1 \times 1)$/Ru(0001) phase and the metallic Ru bulk peak.
In contrast, emission from the Ru${}^{\rm cus, 5f}$ atoms in the
RuO${}_2$(110)-O${}^{\rm cus}$ termination will lead to the
well separated satellite peak discussed in the previous section,
cf. Fig. \ref{sclstrend}c, therewith offering a fingerprint
for the existence of this domain type on the surface. With the presence
of oxides on the surface the Ru $3d$ XPS spectrum will therefore clearly
be shifted towards higher binding energies as more spectral weight is
transferred to the oxidic peaks. Yet, even with a high-resolution X-ray
source it will be very difficult to distinguish between coexisting
RuO${}_2$(110)-O${}^{\rm bridge}$ and O$(1 \times 1)$/Ru(0001) domains.

This distinction will be much more clearcut in the O $1s$ XPS spectrum.
The peak due to RuO${}_2$ lattice oxygen appears at -0.64 eV towards
lower binding energies, so that the largely shifted surface peaks of both
oxide terminations show a total displacement of $\sim$ -1.5 eV. While
this will thus not allow to determine which surface termination is
present on the oxidic domains, the existence of the latter on the
surface (even to a very small degree) will be easily monitored by the
appearance of these new, well-separated peaks in the O $1s$ spectrum.
Together with the Ru $3d$ satellite peak as a fingerprint for the
high-pressure termination, all three domains should thus be clearly
distinguishable by XPS, confirming our initial statement that the latter
technique is a well suited tool to study the initial stages of the
oxidation of the Ru(0001) surface.

\section{Summary}

The oxidation of the Ru(0001) surface proceeds via the O$(1 \times 1)$
chemisorption phase before RuO${}_2$(110) domains start to form. From
a calculation of the corresponding SCLSs we predict a shift of the
Ru $3d$ (O $1s$) XPS spectrum towards higher (lower) binding energies
as soon as oxide domains are present. In particular O $1s$ XPS will
be a very sensitive tool to study the onset of the oxidation process,
as the corresponding large shift results in the appearance of new peaks.
The Ru $3d$ spectrum of RuO${}_2$(110) exhibits a clearly
separated satellite peak at the higher binding energy side, which
we assign to the existence of a hitherto unaccounted high-pressure
termination. In this termination, terminal O${}^{\rm cus}$ atoms sit
atop of the formerly coordinatively unsaturated (cus) Ru${}^{\rm cus, 5f}$
atoms of the stoichiometric RuO${}_2$(110) termination, normally believed to be
the most stable truncation of all rutile (110) crystals. The very
short bondlength between the O${}^{\rm cus}$ and Ru${}^{\rm cus, 6f}$
atoms significantly deviates from the bulk-like situation, leading
to a large initial-state shift of the core-level of the latter
and thus possibly causing the long disputed satellite peak.

\section{Acknowledgements}

Stimulating discussions with Artur B\"ottcher and Horst Conrad
are gratefully acknowledged.

\end{document}